# PROGRESS ON AN ELECTRON BEAM PROFILE MONITOR AT THE FERMILAB MAIN INJECTOR*

R. Thurman-Keup†, T. Folan, M. Mwaniki, S. Sas-Pawlik,
Fermi National Accelerator Laboratory, Batavia, USA

## Abstract

The current program at Fermilab involves the construction of a new superconducting linear accelerator (LINAC) to replace the existing warm version. The new LINAC, together with other planned improvements, is in support of proton beam intensities in the Main Injector (MI) that will exceed 2 MW. Measuring the transverse profiles of these high intensity beams in a ring requires non-invasive techniques. The MI uses ionization profile monitors as its only profile system. An alternative technique involves measuring the deflection of a probe beam of electrons with a trajectory perpendicular to the proton beam. This type of device was installed in MI and initial studies of it have been previously presented. This paper will present the status and recent studies of the device utilizing different techniques.

## INTRODUCTION

Measuring the transverse profile of circulating beam in a proton synchrotron is a challenging endeavor. Historically, thin wires were rapidly swept through the beam and downstream losses recorded as a function of the position of the wire to construct the profile. These were called flying wires [1]. In the Main Injector (MI) at Fermilab, they were removed around 2012 due to wire breakage which was not understood. Since it was already known that wires in the circulating beam might pose a problem at the even higher planned intensities, work had begun on building an electron beam profiler (EBP) to complement the ionization profile monitors that were already installed.

The concept of a probe beam of charged particles to determine a charge distribution has been around since at least the early 1970's [2-4]. Several conceptual and experimental devices have been associated with accelerators around the world [5-9]. An operational device is presently in the accumulator ring at the Spallation Neutron Source at Oak Ridge National Lab [10]. The EBP was constructed and installed in the Main Injector (MI) in 2014 and initial results have been presented previously [11]. This paper presents an update on the status of the EBP.

## THEORY

The principle behind the EBP is electromagnetic deflection of the probe beam by the target beam under study (Fig. 1). If one assumes a target beam with $\gamma \gg 1$, no magnetic field, and $\rho \neq f(z)$, then the force on a probe particle is [12]

$$\vec{F}(\vec{r}) \propto \int d^2\vec{r}' \rho(\vec{r}') \frac{(\vec{r} - \vec{r}')}{|\vec{r} - \vec{r}'|^2} \quad (1)$$

and the change in momentum is

$$\Delta \vec{p} = \int_{-\infty}^{\infty} dt \ \vec{F}(\vec{r}(t)) \quad (2)$$

For small deflections, $\vec{r} \approx \{b, vt\}$, and the change in momentum is

$$\Delta \vec{p} \propto \int_{-\infty}^{\infty} dx' \int_{-\infty}^{\infty} dy' \ \rho(x', y') \\ \cdot \int_{-\infty}^{\infty} dt \ \frac{\{b - x', vt - y'\}}{(b - x')^2 + (vt - y')^2} \quad (3)$$

where {} indicates a vector. For small deflections, $\vec{p} \approx \{0, p\}$ and the deflection is $\theta \approx \frac{|\Delta \vec{p}|}{|\vec{p}|}$. The integral over time can be written as $\text{sgn}(b - x')$ leading to an equation for the deflection

$$\theta(b) \propto \int_{-\infty}^{\infty} dx' \int_{-\infty}^{\infty} dy' \rho(x', y') \, \text{sgn}(b - x') \quad (4)$$

where $\text{sgn}(x) = -1$ for $x < 0$ and $+1$ for $x \geq 0$.

If one takes the derivative of $\theta(b)$ with respect to $b$, the sgn function becomes $\delta(b - x')$ leading to

$$\frac{d\theta(b)}{db} \propto \int_{-\infty}^{\infty} dy' \rho(b, y') \quad (5)$$

which is the profile of the charge distribution of the beam. Thus, for a gaussian beam, this would be a gaussian distribution and the original deflection angle would be the error function, $\text{erf}(b)$.

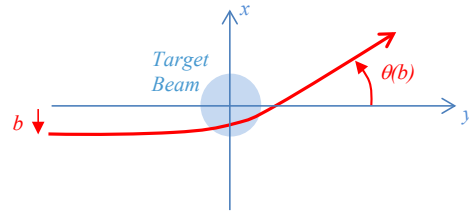

Figure 1: Probe beam deflection (red) for some impact parameter $b$.

## EXPERIMENTAL TECHNIQUE

To obtain $\theta(b)$, one needs to measure the deflection for a range of impact parameters. This can be accomplished by stepping the electron beam through the protons and recording the deflection at each step, or by rapidly sweeping the electron beam through the proton beam (Fig. 2). The latter works without issue provided the sweep time is much smaller than the rms bunch length of the proton beam to avoid coupling the longitudinal and transverse distributions. In the MI, this condition is not met since the bunch length is only 1-2 ns.

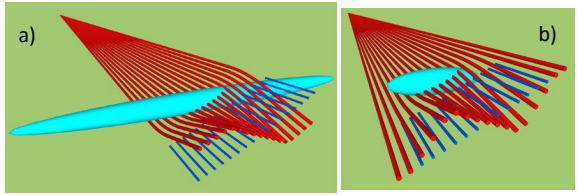

Figure 2: a) Long bunch relative to streak time. b) Short bunch relative to streak time (e.g. MI) leading to bunch-length distortion of deflections.

An alternative method that will be attempted for the MI is a type of raster scan. The beam is swept rapidly along the direction of the proton beam producing an approximate longitudinal distribution (Fig. 3), and slowly transverse to the proton beam. This technique has the potential to allow longitudinal slicing of the transverse profile assuming the longitudinal distribution either remains constant over the series of impact parameter measurements or can be corrected for synchrotron motion.

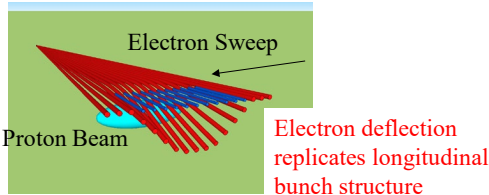

Figure 3: The electron beam is swept along the direction of the proton beam with a sweep time comparable to the proton bunch length. This records the deflection as a function of longitudinal position. A series of these sweeps is collected at different impact parameters to obtain $\theta(b)$.

Simulations of the deflections in this approach are shown in Fig. 4. Here successive sweeps along the proton direction at different impact parameters are displayed in the same image. Each simulated electron produces a Gaussian spot with a rms of 1 mm. The simulation was done for injection parameters of 3 mm horizontal proton beam size, and 2 ns bunch length. One can see that the central deflections may overlap each other.

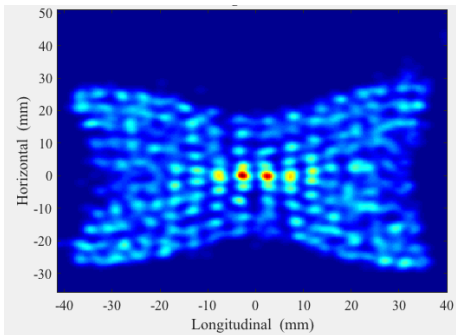

Figure 4: Simulated image of successive electron beam sweeps along the proton beam direction. The sweeps near the center are difficult to separate and may need to be split across multiple camera frames.

Problems such as these must be overcome through, for example, timing shifts or interleaving across multiple camera frames. Before attempting this approach, we have performed a slow diagonal sweep as in Fig. 2 but covering many bunches. The deflections are not as clear, but it is easier to set up.

## EXPERIMENTAL DEVICE

The EBP is comprised of an EGH-6210 electron gun from Kimball Physics, followed by a cylindrical, parallel-plate electrostatic deflector, and finally a phosphor screen acting as the beam dump (Fig. 5).

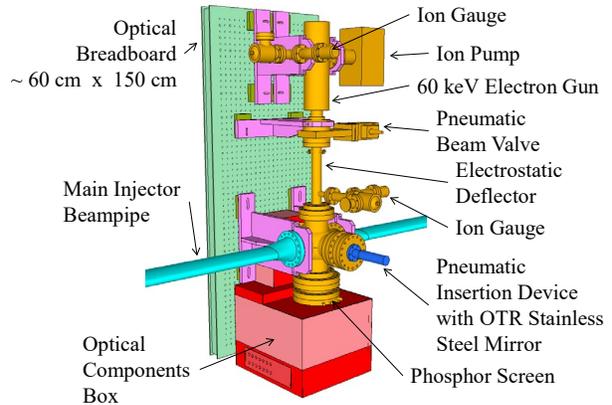

Figure 5: Model of the EBP showing the main components.

The gun is a 60 keV, 6 mA, thermionic gun with a $LaB_6$ cathode, that can be gated from 2 µs to DC at a 1 kHz rate. The gun contains a focusing solenoid and four independent magnet poles for steering/focusing. The minimum working spot size is <100 µm. The electrostatic deflector (Fig. 6) contains 4 cylindrical plates that are 15 cm long and separated by ~2.5 cm. Following the electrostatic deflector is the intersection with the proton beamline.

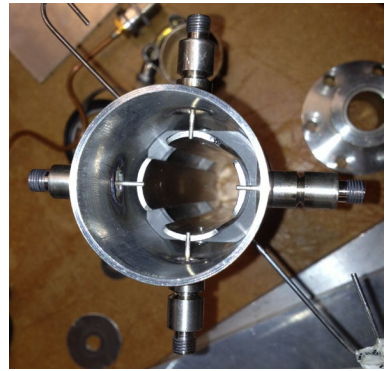

Figure 6: Electrostatic deflector showing the curved plates.

After the proton beam intersection, there is a phosphor screen from Beam Imaging Systems. It is composed of P47 ($Y_2SiO_5$:Ce3+) with an emission wavelength of 400 nm, a decay time of ~60 ns and a quantum yield of 0.055 photons/eV/electron. The phosphor screen has a thin conductive coating with a drain wire attached. The phosphor screen is surrounded by a stainless-steel absorber that is instrumented to function as a faraday cup (Fig. 7).

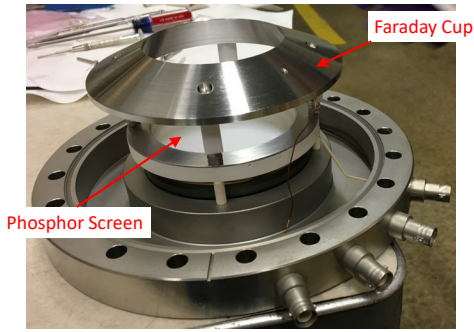

Figure 7: Phosphor screen and angled faraday cup. The faraday cup is intended to be the starting and ending locations for the electron beam during a sweep.

The light from the phosphor traverses a two-lens system plus optional neutral density filters or polarizers before entering a Hamamatsu V6887U-02 gated image intensifier. The output of the intensifier is imaged by a 3710DX12 charge injection device (CID) camera from Spectra-Physics (now Thermo Scientific) which is fiber-optically coupled to the intensifier through a fiber-optic taper to improve light collection (Fig. 8).

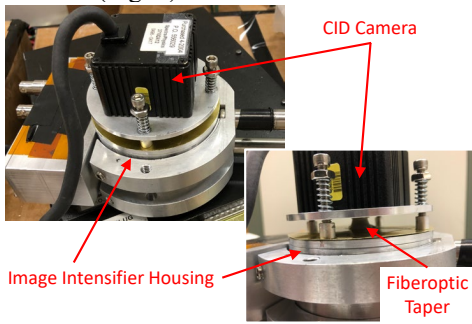

Figure 8: Combination of image intensifier and CID camera coupled together with a fiberoptic taper to map the larger intensifier output to the camera.

Timing for the system is generated by accelerator clock events and requires triggers for camera reset, image acquisition, image intensifier gating, electron gun trigger, fast deflection trigger, and slow deflection trigger. The use and arrangement of the triggers depends on the measurement approach.

## MEASUREMENTS

### Slow Diagonal Sweep

As stated above, we have performed a slow diagonal sweep utilizing adjacent pairs of deflecting plates to sweep the electron beam diagonally across the proton beam. The sweep is relatively slow such that there are several beam revolutions between deflection measurements (Fig. 9).

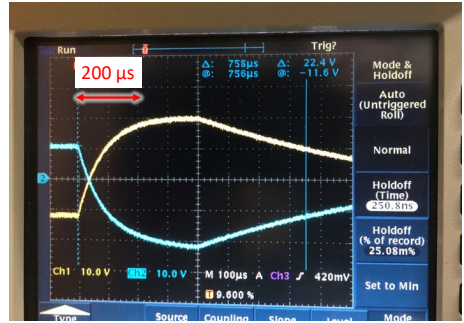

Figure 9: Slow deflection waveforms. A MI revolution is 11 µs.

Figure 10 shows images of the sweep both with and without MI beam. These are an average over 10 camera acquisitions. The individual spots are when the image intensifier is gated. A slight enhancement can be seen in the beam image due to the deflections of the electron beam.

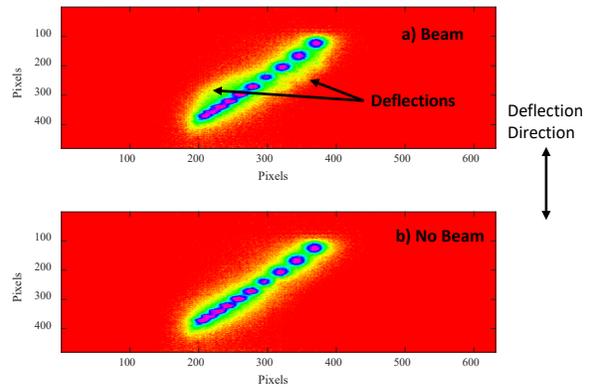

Figure 10: Slow sweep of electron beam. The individual spots are image intensifier gates. A slight enhancement from the electron deflection is visible in a) and labeled.

To extract the deflection, we subtract the no-beam image from the beam image (Fig. 1), where the deflections are now more obvious.

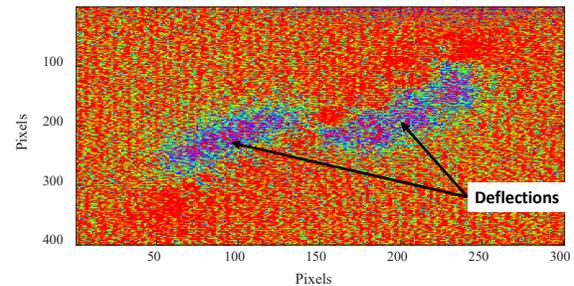

Figure 11: Visible deflections after no-beam image subtraction.

From this image, we find the maximum pixel value in each column. We also find the maximum pixel in each column of Fig. 10b. In Fig. 12, these two curves of maximum pixels are overlayed on the image from Fig. 11.

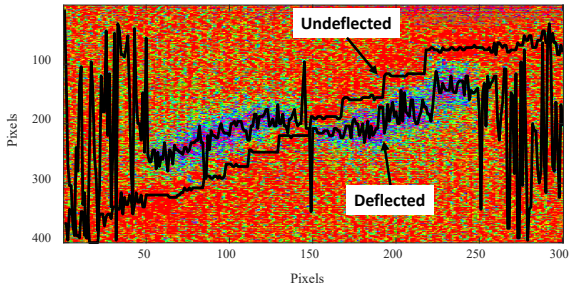

Figure 12: Dark lines show the maximum pixel location for both the deflected electrons and the undeflected electrons.

The undeflected curve shows the step structure of the discrete spots in the original image (Fig. 10b). The difference between these two curves is the deflection distance. We evaluate the average deflection distance for each step region, together with the impact parameter for each step to produce curve (b) in Fig. 13.

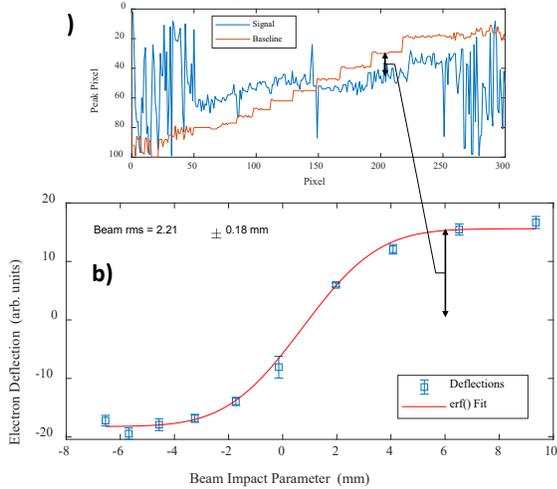

Figure 13: a) The two maximum pixel curves. b) The deflection vs. impact parameter curve generated from (a). The resulting fit sigma is 2.21 mm.

The curve is fit with an error function to extract the underlying gaussian parameters of the proton beam. The sigma of the fit is $2.21 \pm 0.18$(stat.) mm. This is in reasonable agreement with the same measurement from the ionization profile monitor, 1.71 mm.

It is not clear that using the peak pixel is the best approach, and we may refine this technique as we proceed.

## Fast Sweep

The original goal of the EBP was to perform the fast sweep shown above in Fig. 3. Figure 14 shows a scope trace of the fast deflection waveform. We have made one test of the fast sweep (albeit without an actual measurement) and it is shown in Fig. 15. The size of the sweep line is larger than naively expected. This will make the measurement more difficult as we may not be able to put more than one sweep in a single image.

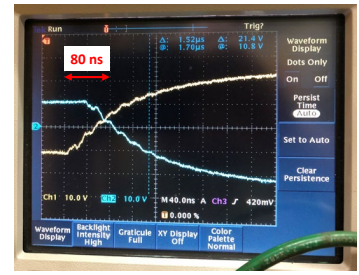

Figure 14: Fast deflection waveforms. A single rf bucket is ~18.8 ns.

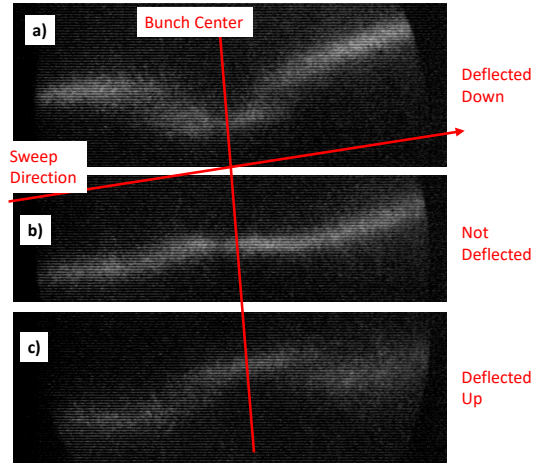

Figure 15: Fast sweep images for the electron beam: a) above the proton beam, b) through the proton beam, and c) below the proton beam.

## CONCLUSIONS

We have successfully measured the proton beam profile with a single camera image using a slow sweep method. It is in reasonable agreement with the same measurement from the ionization profile monitor. The fast sweep method has been attempted and some initial traces were shown. More work is required on this approach before we can make a measurement.